\documentclass[11pt]{article}

\let\fn\footnote
\renewcommand{\footnote}[1]{\linespread{1.1}\fn{#1}\linespread{1.29}}

\makeatletter\renewcommand{\section}{\@startsection {section}{1}{\z@}{-3.5ex plus -1ex minus     -.2ex}{2.3ex plus .2ex}{\bf }}
\makeatletter\renewcommand{\subsection}{\@startsection{subsection}{2}{\z@}{-3.25ex plus -1ex minus    -.2ex}{1.5ex plus .2ex}{\it }}
\makeatletter\renewcommand{\subsubsection}{\@startsection{subsubsection}{3}{-2.45ex}{-3.25ex plus -1ex minus -.2ex}{1.5ex plus .2ex}{\it }}
\renewcommand{\thesection}{\arabic{section}.}
\renewcommand{\thesubsection}{\arabic{section}.\arabic{subsection}.}

\renewcommand{\theequation}{\thesection\arabic{equation}}
\makeatletter \@addtoreset{equation}{section}

\renewenvironment{thebibliography}[1]
     {\baselineskip=16pt plus 2pt minus 1pt
      \section*{\large\refname
        \@mkboth{\MakeUppercase\refname}{\MakeUppercase\refname}}%
     \list{\@biblabel{\@arabic\c@enumiv}}%
           {\settowidth\labelwidth{\@biblabel{#1}}%
            \leftmargin\labelwidth
            \advance\leftmargin\labelsep
            \@openbib@code
            \usecounter{enumiv}%
            \let\p@enumiv\@empty
            \renewcommand\theenumiv{\@arabic\c@enumiv}}%
      \sloppy
      \clubpenalty4000
      \@clubpenalty \clubpenalty
      \widowpenalty4000%
      \sfcode`\.\@m}

\newcommand{\acknowledgements}{\section*{Acknowledgements}
\addcontentsline{toc}{section}{\hspace{0.6cm}{\bf Acknowledgements}}}

\newcommand{\appendices}{\section*{Appendix}\setcounter{subsection}{0}\setcounter{equation}{0}\renewcommand{\thesubsection}{\Alph{subsection}.}
\renewcommand{\theequation}{\thesubsection\arabic{equation}}
\addtocontents{toc}{\vspace{0.2cm}

{\bf Appendices}} }

\usepackage{epsfig,rotating}
\usepackage{amsmath,amssymb}
\usepackage{amsfonts}
\usepackage{mathrsfs}
\usepackage{bbm}
\usepackage{bm}

\def\slasha#1{\setbox0=\hbox{$#1$}#1\hskip-\wd0\hbox to\wd0{\hss\sl/\/\hss}}

\def\periodb#1{\setbox0=\hbox{$#1$}#1\hskip-\wd0\hbox to\wd0{-}}

\newcommand{\CA}{\mathcal{A}}    			

\newcommand{\CC}{\mathcal{C}}

\newcommand{\CD}{\mathcal{D}}

\newcommand{\CL}{\mathcal{L}}

\newcommand{\CN}{\mathcal{N}}
\newcommand{\CO}{\mathcal{O}}

\newcommand{\CR}{\mathcal{R}}

\newcommand{\CV}{\mathcal{V}}

\newcommand{\CW}{\mathcal{W}}

\newcommand{\frg}{\mathfrak{g}}				

\newcommand{\FR}{\mathbbm{R}}     			
\newcommand{\FC}{\mathbbm{C}}     			
\newcommand{\FS}{\mathbbm{S}}     			
\newcommand{\CPP}{{\mathbbm{C}P}}    			

\newcommand{\dd}{\mathrm{d}}     			
\newcommand{\dpar}{\partial}     			
\newcommand{\de}{\mathrm{e}}     			
\newcommand{\di}{\mathrm{i}}     			
\newcommand{\eps}{{\varepsilon}}			

\newcommand{\bpsi}{{\bar{\psi}}}
\newcommand{\bth}{{\bar{\theta}}}
\newcommand{\bphi}{{\bar{\phi}}}

\newcommand{\ald}{{\dot{\alpha}}}     			
\newcommand{\bed}{{\dot{\beta}}}

\newcommand{\eand}{{~~~\mbox{and}~~~}}     		

\newcommand{\tr}{\,\mathrm{tr}\,}     			

\newcommand{\sU}{\mathsf{U}}     			

\newcommand{\sSU}{\mathsf{SU}}

\newcommand{\sMat}{\mathsf{Mat}}

\newcommand{\sSO}{\mathsf{SO}}

\newcommand{\remark}[1]{}     				
\def\tyng(#1){\hbox{\tiny$\yng(#1)$}}			
\def\tyoung(#1){\hbox{\tiny$\young(#1)$}}			

\begin{document}
\begin{titlepage}
\begin{flushright}
  TCDMATH 08-19\\
  HMI 08-03
\end{flushright}
\vskip 2.0cm
\begin{center}
{\LARGE \bf On Superspace Actions for Multiple M2-Branes,\\[0.5cm] 
Metric 3-Algebras and their Classification}
\vskip 1.5cm
{\Large Sergey A.\ Cherkis, Vladimir Dotsenko and Christian S{\"a}mann}
\setcounter{footnote}{0}
\renewcommand{\thefootnote}{\arabic{thefootnote}}
\vskip 1cm
{\em Hamilton Mathematics Institute and\\
School of Mathematics,\\
Trinity College, Dublin 2, Ireland}\\[5mm]

{E-mail: {\ttfamily cherkis, vdots, saemann@maths.tcd.ie}} 
\vskip 1.1cm
\end{center}
\vskip 1.0cm
\begin{center}
{\bf Abstract}
\end{center}
\begin{quote}
We classify canonical metric 3-algebra structures on matrix algebras and find novel three-dimensional conformally invariant actions in ${\CN}=4$ projective superspace based on them.  These can be viewed as Chern-Simons theories with special matter content and couplings. We explore the relations of these actions with the ${\CN}=2$ actions based on generalized 3-Lie algebras found earlier and relate them to the original Bagger-Lambert-Gustavsson action.
\end{quote}
\end{titlepage}

\section{Introduction}

The effective description of stacks of M2-branes which has been proposed by Bagger, Lambert, and Gustavsson in \cite{Bagger:2006sk,Bagger:2007jr,Bagger:2007vi,Gustavsson:2007vu} is based on so-called 3-Lie algebras. These algebras had been introduced in \cite{Filippov:1985aa} as a generalization of the notion of a Lie algebra: a 3-Lie algebra is characterized by a totally antisymmetric trilinear product satisfying a Jacobi-type identity. It was soon realized, however, that if one demands the positivity of the kinetic terms in the action there remains an essentially unique such 3-algebra. Other notions of 3-algebras were explored in \cite{Bagger:2008se} and \cite{Cherkis:2008qr} and, respectively, $\CN=6$ and $\CN=2$ actions based on them were constructed.  These 3-algebras were put into a Lie-theoretic framework in \cite{deMedeiros:2008zh}, where it was established that they are in one-to-one correspondence with pairs $(\frg,V)$ of a metric Lie algebra $\frg$ and its faithful unitary representation $V$.  This analysis also showed that there are two kinds of relevant 3-algebras: real \cite{Cherkis:2008qr} and Hermitian \cite{Bagger:2008se}, depending on whether the representation $V$ is real or complex.

If one is to describe effectively such BLG-like theories via the AdS/CFT correspondence, one would seek a formulation allowing for arbitrary `rank' $N$; in particular, one would like to explore the $N\rightarrow\infty$ limit. With this purpose in mind, we classify all 3-brackets on matrix algebras (or, more generally, any $*$-algebras) that give rise to metric real or Hermitian 3-algebras. These brackets are constructed using only the matrix product and an involution.  Their structure is functorial, i.e.\ it is independent of the rank of the matrices.  This also lists all possible candidates for a notion of a representation for 3-algebras. Recall that in the case of Lie algebras, a representation is a homomorphism into a matrix algebra which takes the Lie bracket into the commutator. For the case of 3-algebras, the matrix product is not enough to produce 3-brackets on matrix algebras which are general enough. However, we expect that considering matrix algebras as $*$-algebras may give enough freedom to represent any 3-algebra.

Besides classifying the 3-brackets, we formulate a manifestly $\CN=4$ supersymmetric action in projective superspace, which is based on these metric 3-algebras\footnote{For the harmonic superspace formulation of the ABJM model \cite{Aharony:2008ug} see  \cite{Buchbinder:2008vi}. This formulation makes $\CN=3$ supersymmetries manifest.}.  We explore their relation with the $\CN=2$ actions found in \cite{Cherkis:2008qr} and with the original action of \cite{Bagger:2007jr}. One of the most interesting features of the BLG model is the subtle interplay between the gauge algebra and supersymmetry, and we hope that our  manifestly supersymmetric formulations shed more light on this issue.

This paper is organized as follows. We start with reviewing the BLG model and the algebraic structure underlying this theory in Section 2. Section 3 contains the classification \eqref{eqA} and \eqref{eqB} of functorial metric 3-algebras structures on $*$-algebras. Section 4 relates the $\CN=2$ actions based on real generalized 3-Lie algebras of \cite{Cherkis:2008qr} with the action of Section 2.  In Section 5, we formulate the $\CN=4$ action in projective superspace for any 3-algebra, whether real or Hermitian. The last section contains our conclusions.

\section{Review of the BLG theory}

If not otherwise specified, we will use the conventions of \cite{Cherkis:2008qr} in the following.

\subsection{Algebraic structure}

Let $\CA$ be a real finite-dimensional metric 3-Lie algebra. That is, $\CA$ is a real vector space endowed with a totally antisymmetric 3-bracket $[\cdot,\cdot,\cdot]:\CA\times\CA\times\CA\rightarrow \CA$ which satisfies the fundamental identity
\begin{equation}
 [A,B,[C,D,E]]\ =\ [[A,B,C],D,E]+[C,[A,B,D],E]+[C,D,[A,B,E]]~,
\end{equation}
and a positive definite bilinear symmetric map $(\cdot,\cdot)_{\CA}:\CA\times\CA\rightarrow \FR$ satisfying the compatibility condition 
\begin{equation}
 ([A,B,C],D)_\CA+(C,[A,B,D])_\CA\ =\ 0~,
\end{equation}
where $A,B,C,D,E\in\CA$. Assuming that $\CA$ is spanned by\footnote{Contrary to common conventions, we follow \cite{Bagger:2007jr} and associate upper indices with the generators of $\CA$.} $\tau^a$, $a,=1,\ldots ,\dim \CA$, we can introduce metric coefficients and structure constants according to
\begin{equation}
h^{ab}_\CA\ =\ (\tau^a,\tau^b)_\CA \eand f^{abcd}_\CA\ =\ f_\CA^{abc}{}_eh^{de}_\CA\ =\ ([\tau^a,\tau^b,\tau^c],\tau^d)_\CA~,
\end{equation}
respectively. Note that the symbol $f^{abcd}_\CA$ is totally antisymmetric. Basically, the only nontrivial example \cite{Nagy:2007aa,Papadopoulos:2008sk,Gauntlett:2008uf} is the metrized version of the 3-Lie algebra $\CA=A_4$  in the classification of \cite{Filippov:1985aa}, for which $\dim \CA=4$, $f^{abcd}_\CA=\eps^{abcd}$; it can be trivially extended to a metric 3-Lie algebra by putting $h^{ab}_\CA=\delta^{ab}$. The only other possibilities are direct sums of $A_4$.

There is an associated Lie algebra $\frg(\CA)$ obtained from $\CA$. This algebra is the linear span of all operators\footnote{We define the wedge and the triple bracket as acting solely on the 3-algebra part. That is, for fields $A,B,C$ taking values in $\CA$, we have $A\wedge B:=A_aB_b \tau^a\wedge\tau^b$ and $[A,B,C]:=A_aB_bC_c[\tau^a,\tau^b,\tau^c]$.} $D_{\tau^a\wedge\tau^b}$ defined by 
\begin{equation}
 D_{\tau^a\wedge\tau^b}(A)\ :=\ [\tau^a,\tau^b,A]~,~~~A\in\CA~.
\end{equation}
Thus, the space $\CA$ is a carrier space of a faithful representation of $\frg(\CA)\subseteq \mathfrak{o}(\CA)$. The Lie algebra $\frg(\CA)$
has structure constants\footnote{In the following, subscripts $\frg$ refer to $\frg(\CA)$ for short.}
\begin{equation}
f_{\frg}^{[ab][cd]}{}_{[ef]}\ :=\ (f_\CA^{abc}{}_e\delta^d_f+\delta^c_ef_\CA^{abd}{}_f)_{[ef]}~.
\end{equation}
Here, $(\cdot)_{[ef]}$ denotes antisymmetrization in the indices $e,f$. The induced Killing form on $\frg(\CA)$ is given by
\begin{equation*}
(D_{\tau^a\wedge\tau^b},D_{\tau^c\wedge\tau^d})_{\frg,\mathrm{ind}}\sim \sum_{[ef],[gh]} f_\frg^{[ab][ef]}{}_{[gh]}f_\frg^{[cd][gh]}{}_{[ef]}=:h^{[ab][cd]}_{\frg,\mathrm{ind}}~,
\end{equation*}
and direct computation\footnote{Note that $f_\CA^{abc}{}_c=f^{abcd}_\CA h^\CA_{cd}=0$, where $h^\CA_{ab}$ is the inverse of $h^{ab}_\CA$.} shows that 
\begin{equation}
 h^{[ab][cd]}_{\frg,\mathrm{ind}}\ =\ (\dim\CA-2)f_\CA^{abe}{}_gf_\CA^{cdg}{}_e~.
\end{equation}
Whenever the Cartan criterion $h^{[ab][cd]}_{\frg,\mathrm{ind}}\sim \delta^{[ab][cd]}$ is fulfilled (which is the case e.g.\ for the 3-Lie algebra $A_4$), $\frg(\CA)$ is a semisimple Lie algebra.

For the BLG theory, however, one has to work with another pairing $(\cdot,\cdot)_\frg$ with coefficients
\begin{equation}
(D_{\tau^a\wedge\tau^b},D_{\tau^c\wedge\tau^d})_{\frg}\ =\ f_\CA^{abcd}=:h^{[ab][cd]}_\frg~.
\end{equation}
The latter choice is symmetric and satisfies in particular $([X,Y],Z)_\frg+(Y,[X,Z])_\frg=0$ for all $X,Y,Z\in\frg(\CA)$ and is, therefore, an invariant pairing. It is also non-degenerate, as the original pairing on $\CA$ is. It has been proved in \cite{deMedeiros:2008zh} that $h_{\frg}^{[ab][cd]}$  always has split signature and therefore necessarily differs from the Killing form. It follows that $\frg(\CA)$ is not a simple Lie algebra.

\subsection{Action, equations of motions, and supersymmetry transformations}

The matter sector of the theory is given by 8 scalars $X^I$, $I=1,\ldots ,8$ together with a Majorana spinor $\Psi$ of $\sSO(1,10)$, both taking values in the 3-Lie algebra $\CA$. Additionally, we introduce a gauge potential $A_\mu$, which takes values in the associated Lie algebra $\frg(\CA)$ and yields the covariant derivative
\begin{equation}
 \nabla_\mu X\ =\ \dpar_\mu X-A_\mu X~.
\end{equation}
Infinitesimal gauge transformations generated by $\Lambda\in\frg(\CA)$ act according to
\begin{equation} \label{gaugetrafos}
 \delta X\ =\ \Lambda X~,~~~\delta\Psi\ =\ \Lambda \Psi~,~~~\delta A_\mu\ =\ \dpar_\mu\Lambda-[A_\mu,\Lambda]~.
\end{equation}
Let $\hat{\Gamma}_M$, $M=0,\ldots,10$, be the $32\times 32$-dimensional $\gamma$-matrices generating the Clifford algebra $\CC:=Cl(\FR^{1,10})$ and $\Gamma:=\hat{\Gamma}_0\ldots.\hat{\Gamma}_{10}$ be the corresponding chirality operator; introduce furthermore the subset $\Gamma_I=\hat{\Gamma}_{I+2}$, $I=1,...,8$. We find it convenient to combine the scalars into $X:=\Gamma_I X^I$, and to extend the definition of the triple bracket linearly:
\begin{equation}
 [X,X,A]:=\Gamma_{IJ}[X^I,X^J,A]\eand[X,X,X]:=\Gamma_{IJK}[X^I,X^J,X^K]~,
\end{equation}
where $A\in\CA$ and $\Gamma_{I\ldots J}=\Gamma_{[I}\ldots\Gamma_{J]}$. Moreover, we extend the trace over $\CA$ to a trace over $\CA\otimes \CC$ by
\begin{equation}
 (A,B)_{\CA\otimes \CC}:=\tfrac{1}{32}\tr_\CC\big( (A,B)_\CA\big)~,
\end{equation}
where $\tr_\CC(\cdot)$ denotes the trace over the $\gamma$-matrices. With these definitions, we can write the Lagrangian of the BLG model \cite{Bagger:2007jr} as 
\begin{equation} \label{BLG-theory}
\begin{aligned}
 \CL_{\mathrm{BLG}}\ =\ &-\tfrac{1}{2}(\nabla_\mu X,\nabla^\mu X)_{\CA\otimes\CC}+\tfrac{\di}{2}(\bar{\Psi},\Gamma^\mu \nabla_\mu\Psi)_\CA +\tfrac{\di}{4}(\bar{\Psi},[X,X,\Psi])_\CA\\
&-\tfrac{1}{12}([X,X,X],[X,X,X])_{\CA\otimes\CC} +\tfrac{1}{2}\eps^{\mu\nu\kappa}\left((A_{\mu},\dpar_\nu A_{\kappa})_\frg+\tfrac{1}{3}(A_{\mu},[A_{\nu},A_{\kappa}])_\frg\right)~.
\end{aligned}
\end{equation}
The associated action is invariant (up to a total derivative) under the gauge transformations \eqref{gaugetrafos}. It is furthermore invariant under $\CN=8$ supersymmetries up to the equations of motion
\begin{equation} \label{BLG-eom}
\begin{aligned}
  \nabla_\mu \nabla^\mu X+\tfrac{\di}{2}\Gamma_K[\bar{\Psi}, X,\Gamma^K\Psi]+\tfrac{1}{2} \Gamma[X,X,\Gamma[X, X, X]]&\ =\ 0~,\\
 \Gamma^\mu \nabla_\mu \Psi+\tfrac{1}{2}[X,X,\Psi]&\ =\ 0~,\\
[\nabla_\mu,\nabla_\nu]+\eps_{\mu\nu\kappa}(\tr_\CC(X\wedge(\nabla^\kappa X))+\tfrac{\di}{2}\bar{\Psi}\wedge(\Gamma^\kappa \Psi))&\ =\ 0~.
\end{aligned}
\end{equation}
Explicitly, the supersymmetry transformations are given by
\begin{equation}
  \delta X \ =\  \di \Gamma_I\bar{\eps}\Gamma^I\Psi~,~~~
  \delta \Psi\ =\ \nabla_\mu X\Gamma^\mu\eps-\tfrac{1}{6} [X,X,X]\eps~,~~~
  \delta A_\mu\ =\ \di\bar{\eps}\Gamma_\mu (X\wedge\Psi)~.
\end{equation}

\section{Generalizations and the classification of 3-algebras}

In \cite{deMedeiros:2008zh}, it was observed that 3-Lie algebras and two generalizations introduced in \cite{Bagger:2008se} and \cite{Cherkis:2008qr} can be constructed by a procedure originally introduced by Faulkner \cite{Faulkner:1973aa}. Namely, one starts with a pair $(\frg,V)$ consisting of a metric Lie algebra $\frg$ and its faithful unitary representation $V$. From this data, one produces a 3-algebra structure. In fact, there is equivalence of 3-algebras with such pairs up to isomorphisms. The analysis of \cite{deMedeiros:2008zh} yielded three such types\footnote{Some specific 3-algebras appear in \cite{AliAkbari:2008rm} and \cite{SheikhJabbari:2008wq}. These are based on a matrix algebra with a 4-bracket $[[A_1,A_2,A_3,A_4]]:=\tfrac{1}{4!}\eps^{ijkl}A_iA_jA_kA_l$ with a 3-algebra structure defined by $[A,B,C]=[A,B,C,T]$ with an appropriately chosen matrix $T$. This 3-bracket does not necessarily satisfy the fundamental identity. However, in \cite{AliAkbari:2008rm} it is observed that its factor algebra is a 3-Lie algebra, which is either $A_4$ or Lorentzian. In \cite{SheikhJabbari:2008wq}, on the other hand, only part of the associated Lie algebra is gauged and as a result, only some of the fundamental identity relations have to be enforced.}:
\begin{itemize}
 \item[(1)] The real case, obtained from a Lie algebra $\frg$ together with a faithful real orthogonal representation of $\frg$. This yields the generalized 3-Lie algebras of \cite{Cherkis:2008qr}.
 \item[(2)] The Hermitian case, which corresponds to a pair of a Lie algebra and a faithful complex unitary representation. This corresponds to the complex 3-Lie algebras of \cite{Bagger:2008se}.
 \item[(3)] The quaternionic case, in which the representation is quaternionic unitary. This case can be considered as a specialization of the Hermitian case.
\end{itemize}

In this section we present a classification for functorial generalized 3-Lie structures on $*$-algebras. Let us be precise about the problem we are solving.

Recall that a $*$-algebra (over complex numbers) is an associative algebra $\CR$ with antilinear involutive antiautomorphism, that is a map $*\colon \CR\to \CR$, $A\mapsto A^*$ satisfying for all $A,B\in \CR$ and $\alpha\in\mathbb{C}$
\begin{gather}
(\alpha A)^*=\overline{\alpha}\,A^*~,~~~ (A^*)^*=A~,\\
(A+B)^*=A^*+B^*~,~~~ (AB)^*=B^* A^*~.
\end{gather}
We consider algebras equipped with the trace form $\tr\colon \CR\to\mathbb{C}$ that satisfies 
\begin{gather}
\tr(AB)=\tr(BA)~,~~~ \tr(A^*)=\overline{\tr(A)}~,\label{trace}\\
 0\le\tr(AA^*)\in\mathbb{R}~.
\end{gather}
In the case of real $*$-algebras, the definition is the same, but the coefficients $\alpha$ above are real.

The brackets classified below are all functorial brackets, that is brackets that can be defined in terms of the basic operations: the product and the involution. Note that since the bracket has to be multilinear, it should be equal to a combination of cubic monomials. Even though for matrices of small size there may exist brackets outside our classification, this classification is exhaustive for matrices of generic size since these do not satisfy any $*$-algebra identities. More formally, one can say\footnote{If one only cares about the fundamental identity; trace relations specific for matrices of given size make everything a bit more subtle.} that we are looking for homomorphisms from the operad of generalized 3-Lie algebras to the operad of $*$-algebras. In the case of a matrix algebra $\CR=\sMat_n$ with some canonical 3-bracket on it, we are essentially discussing what are $n$-dimensional representations for arbitrary 3-algebras, where the representation homomorphism takes the 3-algebra bracket into the canonical one. 

\subsection{The real case}

Generalized 3-Lie algebras were introduced in \cite{Cherkis:2008qr} and were used to obtain a BLG-type model with $\CN=2$ supersymmetries. Given a generalized 3-Lie algebra, we have a trilinear triple product $[\cdot,\cdot,\cdot]:\CA\times\CA\times\CA\rightarrow\CA$ and a Euclidean form $(\cdot,\cdot)_\CA:\CA\times\CA\rightarrow \FR$ on a real vector space $\CA$, which satisfy the following compatibility conditions for all $A,B,C,D,E\in\CA$:
\begin{itemize}
 \item[(1)] The 3-product is antisymmetric in its first two slots:
\begin{equation}
 [A,B,C]\ =\ -[B,A,C]~.
\end{equation}
 \item[(2)] The 3-product satisfies the fundamental identity:
\begin{equation}\label{Eq:FundamentalIdentity}
 [A,B,[C,D,E]]\ =\ [[A,B,C],D,E]+[C,[A,B,D],E]+[C,D,[A,B,E]]~.
\end{equation}
 \item[(3)] The pairing $(\cdot,\cdot)_\CA$ is an invariant form under derivations of $\CA$ given by the 3-product:
\begin{equation}\label{Eq:Comp1}
 ([A,B,C],D)+(C,[A,B,D])\ =\ 0~,
\end{equation}
and satisfies the further symmetry relation
\begin{equation}\label{Eq:Comp2}
  ([A,B,C],D)+([C,D,A],B)\ =\ 0~.
\end{equation}
\end{itemize}

Exactly in the same fashion as for 3-Lie algebras, which were discussed in Section 2.1, we can define a Lie algebra $\frg(\CA)\subseteq\mathfrak{o}(\CA)$ associated to the generalized 3-Lie algebra $\CA$, cf.\ \cite{Cherkis:2008qr}. Moreover, it was proved in \cite{deMedeiros:2008zh} that there is a one-to-one correspondence, up to isomorphisms, between generalized 3-Lie algebras and pairs $(\frg,V)$, where $\frg$ is a metric real Lie algebra and $V$ is a faithful orthogonal $\frg$-module.

\subsection{Classification of real metric 3-algebras}

We begin by identifying all brackets that only satisfy the fundamental identity. Let us use the following notation: 
$$
[A,B]=AB-BA~,\quad \{A,B\}=AB+BA~,\quad [A,B]_C=ACB-BCA~.
$$

The following classification is a result of a straightforward computation. For a generic combination of cubic monomials, we wrote down the fundamental identity, extracted the coefficients of all appearing monomials of degree~5, and, using the computer algebra system \texttt{MuPAD}, we solved the corresponding system of equations. The set of solutions of that system provides the following list of ternary brackets  that arise canonically from real $*$-algebras and satisfy the fundamental identity:
\begin{itemize}
 \item[I$_\alpha$]:
$$
A,B,C\mapsto \alpha(\{[A,B],C^*\}+[\{A^*,B\},C]-[\{A,B^*\},C]-\{[A^*,B^*],C^*\})~;
$$
\item[II$_{\alpha,\beta}$]:
$$
A,B,C\mapsto \alpha[[A,B],C]
+\beta([[A^*,B],C]+[[A,B^*],C]+[[A,B],C^*]-[[A^*,B^*],C^*])~;
$$
\item[III$_{\alpha,\beta}$]:
\begin{multline*}
A,B,C\mapsto \alpha([[A,B],C]+[[A^*,B^*],C]+[[A^*,B],C^*]+[[A,B^*],C^*])\\
+\beta([[A,B],C^*]+[[A^*,B],C]+[[A,B^*],C]+[[A^*,B^*],C^*])~;
\end{multline*}
\item[IV$_{\alpha,\beta}$]:
\begin{multline*}
A,B,C\mapsto \alpha(\{[A^*,B^*],C^*\}+\{[A^*,B^*],C\}+\{[A,B^*],C^*\}+\{[A^*,B],C^*\}\\
+\{[A^*,B],C\}+\{[A,B^*],C\}+\{[A,B],C^*\}+\{[A,B],C\})\\
+\beta([A,C]_B+[A,C^*]_B+ [A,C]_{B^*}+ [A^*,C]_B+ [A,C^*]_{B^*}\\ + [A^*,C^*]_B+ [A^*,C]_{B^*}+ [A^*,C^*]_{B^*})~;
\end{multline*}
\item[V$_{\alpha,\beta}$]:
\begin{multline*}
A,B,C\mapsto \alpha(\{[A^*,B^*],C^*\}-\{[A^*,B^*],C\}-\{[A,B^*],C^*\}-\{[A^*,B],C^*\}\\
+\{[A^*,B],C\}+\{[A,B^*],C\}+\{[A,B],C^*\}-\{[A,B],C\})\\
+\beta(-[A,C]_B+ [A,C^*]_B+ [A,C]_{B^*}+ [A^*,C]_B\\ 
-[A,C^*]_{B^*}-[A^*,C^*]_B-[A^*,C]_{B^*}+ [A^*,C^*]_{B^*})\,;
\end{multline*}
\item[VI$_{\alpha,\beta}$]:
\begin{multline*}
A,B,C\mapsto \alpha^2(\{[A^*,B^*],C^*\}+[\{A^*,B\},C]-[\{A,B^*\},C]+\{[A,B],C^*\})\\
+\alpha\beta([AB^*-BA^*,C^*]+(AB+A^*B^*)C) +\beta^2(AB^*-BA^*)C~;
\end{multline*}
\item[VII$_{\alpha,\beta,\gamma}$]:
\begin{multline*}
A,B,C\mapsto \alpha^3(-[[A,B],C]-[[A^*,B^*],C]+\{\{A^*,B\},C\}-\{\{A,B^*\},C\})\\
+\beta\gamma^2((AB^*-BA^*)C)\\+\alpha^2\beta(\{[A,B],C^*\}+\{[A^*,B^*],C^*\}+[\{A^*,B\},C]-[\{A,B^*\},C])\\
+\alpha^2\gamma((BA^*-AB^*)C+C(A^*B-B^*A))\\+\alpha\beta\gamma([A,B]C+[A^*,B^*]C+\{AB^*-BA^*,C^*\})~;
\end{multline*}
\item[VIII$_{\alpha,\beta}$]:
$$
A,B,C\mapsto \alpha(AB^*-BA^*)C+\beta C(A^*B-B^*A)~;
$$
\item[IX$_{\alpha,\beta,\gamma}$]:
$$
A,B,C\mapsto \alpha([[A,B],C]+[[A^*,B^*],C])+\beta[AB^*-BA^*,C]+\gamma[A^*B-B^*A,C]~.
$$
\end{itemize}
Here, the Greek letters $\alpha, \beta, \gamma$ denote arbitrary real parameters.

Using this result, it is not too hard to obtain the list of canonical 3-brackets satisfying all conditions with the inner product $(A,B)=\tr(AB^*)$. Here we use the fact that all identities involving traces follow from the fundamental properties of traces \eqref{trace}. Among the brackets listed above, all  satisfy the fundamental identity \eqref{Eq:FundamentalIdentity}, but not all are compatible with the inner product.  Imposing these compatibility conditions \eqref{Eq:Comp1} and \eqref{Eq:Comp2}, we end up with the following list of canonical generalized 3-Lie algebras arising from real $*$-algebras:
\begin{itemize}
 \item[I$_{\alpha}$]: 
$$
A,B,C\mapsto \alpha([[A^*,B],C]+[[A,B^*],C]+[[A,B],C^*]-[[A^*,B^*],C^*])~;
$$
 \item[II$_{\alpha}$]:
\begin{equation}\label{eqA}
A,B,C\mapsto \alpha([[A,B^*],C]+[[A^*,B],C])~;
\end{equation}
 \item[III$_{\alpha,\beta}$]:
$$
A,B,C\mapsto \alpha(AB^*-BA^*)C+\beta C(A^*B-B^*A)~;
$$
 \item[IV$_{\alpha,\beta}$]: 
\begin{multline*}
A,B,C\mapsto \alpha([[A,B],C]+[[A^*,B^*],C]+[[A^*,B],C^*]+[[A,B^*],C^*])\\
+\beta([[A,B],C^*]+[[A^*,B],C]+[[A,B^*],C]+[[A^*,B^*],C^*])~.
\end{multline*}
\end{itemize}
As above, $\alpha, \beta$ denote again arbitrary real parameters.

At the moment we do not know a complete answer to the question of how these brackets can be represented by each other (in particular, it might be the case that they all can be represented by one of them). Let us give one nontrivial example of such a representation here: The matrix algebra $\sMat_n$ with the bracket of type IV$_{0,1}$ is isomorphic to a 3-subalgebra of the matrix algebra $\sMat_{2n}$ with the bracket of type III$_{1,-1}$; the embedding is given by the formula $a\mapsto\begin{pmatrix}a&a^*\\a^*&a\end{pmatrix}$.

\subsection{The Hermitian case}

This generalization of a 3-Lie algebra was introduced in \cite{Bagger:2008se} to obtain a BLG-type model preserving $\CN=6$ supersymmetries. A metric Hermitian 3-Lie algebra $\CA$ is given by a 3-product $[\cdot,\cdot\,;\,\cdot]:\CA\times\CA\times\CA\rightarrow \CA$, which is linear in its first two slots and antilinear in the third one, and a Hermitian form $(\cdot,\cdot):\CA\times\CA\rightarrow \FC$, which satisfy the following constraints for all $A,B,C,D,E\in\CA$:
\begin{itemize}
 \item[(1)] Antisymmetry in the first two slots:
\begin{equation}
[A,B;C]=-[B,A;C]~.
 \end{equation}
 \item[(2)] Metric compatibility:
\begin{equation}
 ([A,B;C],D)=(B,[C,D;A])~.
\end{equation}
 \item[(3)] The Hermitian fundamental identity:
\begin{equation} \label{hermitianFundamentalIdentity}
 [[C,D;E],A;B]-[[C,A;B],D;E]-[C,[D,A;B];E]+[C,D;[E,B;A]]\ =\ 0~.
\end{equation}
\end{itemize}

We assume again that $\CA$ as a vector space is spanned by $\tau^a$, $a=1,\ldots ,\dim(\CA)$. Then we find an associated Lie algebra $\frg^\FC(\CA)$ which is generated by the following action on elements of $\CA$:
\begin{equation}
 D_{\tau^a\otimes \tau^b}(A)\ :=\ [A,\tau^a;\tau^b]~,~~~A\in\CA~.
\end{equation}
To obtain the associated real Lie algebra $\frg(\CA)\subseteq \mathfrak{o}(\CA)$, one has to antisymmetrize this action, cf.\ \cite{deMedeiros:2008zh}:
\begin{equation}
 D_{\tau^a\wedge \tau^b}(A)\ :=\ [A,\tau^a;\tau^b]-[A,\tau^b;\tau^a]~,~~~A\in\CA~.
\end{equation}
The Hermitian fundamental identity \eqref{hermitianFundamentalIdentity} ensures \cite{deMedeiros:2008zh} that the commutator of two such actions is again an element in $\frg(\CA)$. An invariant form on $\frg(\CA)$ can be defined as the linear extension of
\begin{equation}
 (D_{\tau^a\wedge \tau^b},D_{\tau^c\wedge \tau^d})_\frg\ :=\ \mathrm{Re}([\tau^c,\tau^a;\tau^b],\tau^d)_\CA~.
\end{equation}
From the above constraints (1) and (2), we obtain $([A,B;C],D)=-([A,B;D],C)$ and as $(\cdot,\cdot)_\CA$ is Hermitian, $(\cdot,\cdot)_\frg$ is also Hermitian and satisfies
\begin{equation}
 (X,Y)_\frg\ =\ (Y,X)^*_\frg
\end{equation}
for all $X,Y\in\frg(\CA)$. Invariance of this Hermitian form follows in a straightforward manner from the axioms given above and for all $X,Y,Z\in\frg(\CA)$, we have
\begin{equation}
([X,Y],Z)_\frg+(Y,[X,Z])_\frg\ =\ 0~.
\end{equation}

\subsection{Classification of Hermitian metric 3-algebras}

The linearity/antilinearity condition is very restrictive.  Instead of 48 cubic monomials that occur in a generic combination, we have only 6. This results in just one series of examples. Namely, it is easy to see that the only series of Hermitian metric 3-Lie algebras arising from complex $*$-algebras is
\begin{equation}\label{eqB}
A,B,C\mapsto\alpha(AC^*B-BC^*A)~,
\end{equation}
where $\alpha\in\mathbb{R}$.

It is interesting to note that this only bracket has been already discovered in \cite{Bagger:2008se}; there this bracket on matrix algebras was used to give a 3-algebraic formulation of the ABJM model.

\section{$\CN=2$ superfield formulation}

It is evident that the BLG theory has to allow for a description in terms of $\CN=2$ superfields in three dimensions that is analogous to Chern-Simons matter theories \cite{Zupnik:1988en,Ivanov:1991fn}, see also \cite{Nishino:1991sr}. All $\CN=2$ superfield actions found in \cite{Cherkis:2008qr} however differ from the BLG theory in the interaction terms. As we will show in this section, one can add a superpotential term to this action such that it exactly reproduces the BLG theory. Note that we give here a general superfield action, which can be combined with any of the 3-algebra structures discussed above, whether real or Hermitian. In the following, we adopt the conventions of \cite{Wess:1992cp} and \cite{Cherkis:2008qr}.

\subsection{The superfield action}

Let $\CA$ be an arbitrary 3-algebra and $\frg(\CA)$ its associated Lie algebra. Consider four $\CA$-valued chiral superfields $\Phi^i$, $i=1,\ldots ,4$ together with a $\frg(\CA)$-valued vector superfield $V$ defined on the superspace $\FR^{1,2|4}$. We will use the following superfield expansions, which are obtained by dimensionally reducing the superfield expansions of \cite{Wess:1992cp} from $\FR^{1,3|4}$ to $\FR^{1,2|4}$:
\begin{equation}
\begin{aligned}
 \Phi^i(y)&\ =\ \phi^i(y)+\sqrt{2} \theta \psi^i(y)+\theta^2 F^i(y)~,\\
 V(x)&\ =\ - \theta^\alpha\bth^\ald(\sigma^{\mu}_{\alpha\ald}A_\mu(x)+\di\eps_{\alpha\ald}\sigma(x))+\di\theta^2(\bth\bar{\lambda}(x))-\di\bth^2(\theta\lambda(x))+\tfrac{1}{2}\theta^2\bth^2 D(x)~,
\end{aligned} 
\end{equation}
where $x\in \FR^{1,2}$ and $y$ is the usual coordinate on the chiral subspace $\FR^{1,2|2}_\mathrm{ch}\subset\FR^{1,2|4}$. BLG-type actions with at least $\CN=2$ supersymmetry are obtained from the following base action \cite{Cherkis:2008qr}:
\begin{equation} \label{CSaction1}
 S_0\ =\ \int \dd^3x\int \dd^4\theta ~\kappa \left(\di ( V,(\bar{D}_{\alpha}D^\alpha V))_\frg+\tfrac{2}{3}(V,\{(\bar{D}^\alpha V),(D_\alpha V)\})_\frg\right)+(\bar{\Phi}_i,\de^{2iV}\Phi^i)_\CA~.
\end{equation}
Note that the anticommutator $\{\cdot,\cdot\}$ reduces in components to the commutator in $\frg(\CA)$ and is therefore well defined. This action possesses manifest $\CN=2$ supersymmetry as well as an $\sSU(4)$ flavor symmetry. The interaction terms, however, do not possess the $\sSO(8)$ flavor symmetry, and therefore the action \eqref{CSaction1} is necessarily different from the BLG model. One can supersymmetrically extend $S_0$ by adding superpotential terms of the form
\begin{equation}
 S_1\ =\ \int \dd^3 x\int \dd^2 \theta \CW(\Phi^1,\ldots ,\Phi^4)+c.c.~,
\end{equation}
where $\CW$ is a (holomorphic) polynomial in the chiral matter superfields $\Phi^i$. Classical conformal invariance requires $\CW$ to be  homogeneous of degree 4. Furthermore, the desired flavor symmetry as well as the R-symmetry for $\CN>2$ restricts the form of the polynomial. For example, for flavor symmetry $\sSU(4)$ and $\CA$ a 3-Lie algebra, the only possible polynomial is
\begin{equation} \label{superpotential}
 \CW(\Phi^1,\ldots ,\Phi^4)\ =\ t_{ijkl}([\Phi^i,\Phi^j,\Phi^k],\Phi^l)_\CA~,
\end{equation}
where $t_{ijkl}$ has to be an $\sSU(4)$ invariant tensors and is therefore proportional to $\eps_{ijkl}$. One can also consider an additional Yang-Mills term; however, such a term does break classical conformal invariance.

\subsection{The BLG model in $\CN=2$ language: The 3-Lie algebra case}

Not surprisingly, and similarly to the description of $d=4$, $\CN=4$ super Yang-Mills theory in terms of $\CN=1$ superfields, one can couple the superpotential \eqref{superpotential} to the base action $S_0$ and for a specific value of the coupling, one obtains the BLG model. For this, we have to demand that $\CA$ is a 3-Lie algebra. Then\footnote{In \cite{Cherkis:2008qr}, it was erroneously stated that this superpotential term is $\sSO(8)$-invariant in general. This is only true in a special situation described in \cite{Bagger:2006sk} and is not the case here.}
\begin{equation}
 S_1\ =\ \int \dd^3 x\int \dd^2 \theta ~\alpha\,\eps_{ijkl}([\Phi^i,\Phi^j,\Phi^k],\Phi^l)_\CA+c.c.~,
\end{equation}
and the choice $\alpha=\frac{\di}{4!\kappa}$ yields, after integrating out all auxiliary fields, the bosonic potential terms
\begin{equation}\label{bospot}
  -\tfrac{1}{6\kappa^2}([\phi^i,\phi^j,\phi^k],[\bphi^i,\bphi^j,\bphi^k])_\CA+\tfrac{1}{4\kappa^2}([\phi^k,\bphi^k,\phi^i],[\phi^j,\bphi^j,\bphi^i])_\CA~.
\end{equation}
Rewritten in terms of the real fields $X^I$, $I=1,\ldots ,8$, defined as $\phi^i:=X^{2i-1}+\di X^{2i}$, the bosonic interaction terms \eqref{bospot} equal to
\begin{equation}
V_{\mathrm{bos}}\ =\  -\tfrac{1}{6\kappa^2}([X^I,X^J,X^K],[X^I,X^J,X^K])_\CA~.
\end{equation}
In the following, we keep $\alpha$ arbitrary, however. For the auxiliary fields $F^i,\sigma,\lambda,D$ we find the equations of motion
\begin{equation}
\begin{aligned}
 F^i&\ =\ -4\bar{\alpha}\eps^{jkli}[\phi^j,\phi^k,\phi^l]~,\\
 \sigma&\ =\ \tfrac{\di}{2\kappa}\phi^i\wedge\bar{\phi^i}~,\\
 \lambda&\ =\ ~\tfrac{1}{\sqrt{2}\di\kappa}\phi^i\wedge\bar{\psi}^i,\\
 D&\ =\ \tfrac{\di}{4\kappa}\left([\phi^j,\bphi^j,\phi^i]\wedge\bphi^i-\phi^i\wedge[\phi^j,\bphi^j,\bphi^i]\right)+\tfrac{1}{2\kappa}\bar{\psi}^{i\alpha}\wedge\psi^i_\alpha~,\\
\end{aligned}
\end{equation}
and eliminating these fields, we arrive at the following Lagrangian in components:
\begin{equation} \label{componentaction}
\begin{aligned}
\CL\ =\ &\kappa\eps^{\mu\nu\kappa}\left((A_\mu,\dpar_\nu A_\kappa)_\frg+\tfrac{1}{3}(A_\mu,[A_\nu,A_\kappa])_\frg\right)-(\nabla_\mu \bphi^i,\nabla^\mu \phi^i)_\CA-\di(\bar{\psi}^i,\sigma^\mu\nabla_\mu \psi^i)_\CA\\
&+\frac{1}{4\kappa^2}\left([\bphi^i,\phi^i,\phi^j],[\bphi^k,\phi^k,\bphi^j]\right)_\CA+\frac{\di}{2\kappa}\left(\bpsi^{j\alpha},[\phi^i,\bphi^i,\psi^j_\alpha]\right)_\CA\\
&+\frac{\di}{\kappa}\left(\bphi^i,[\phi^j,\bpsi^{j\alpha},\psi^i_\alpha]\right)_\CA -96\alpha^2([\phi^i,\phi^j,\phi^k],[\bphi^i,\bphi^j,\bphi^k])_\CA\\
&-2\eps_{ijkl}\left(\alpha([\psi^{i\alpha},\psi^j_\alpha,\phi^k],\phi^l)_\CA+\bar{\alpha}([\bpsi^i_\alpha,\bpsi^{j \alpha},\bphi^k],\bphi^l)_\CA\right)~.
\end{aligned}
\end{equation}

We will now discuss the supersymmetry transformations in more detail. For the manifest $\CN=2$ supersymmetries, explicit expressions are obtained from the de Wit-Freedman transformation, see appendix A. Here, one considers the supervector field in Wess-Zumino gauge and, to preserve this gauge, every supersymmetry transformation needs to be accompanied by a gauge transformation. This makes the transformation law non-linear. 

The supersymmetry transformations act on a superfield $F$ according to
\begin{equation}
 \delta F\ :=\  (\bar{\eps}\bar{Q}+\eps Q) F~.
\end{equation}
The various superfield components transform according to
\begin{equation}
\begin{aligned}
  \delta A_{\mu} &\ =\ \tfrac{1}{\sqrt{2}}(\eps^\alpha\sigma_{\mu \alpha\beta}\psi^{i\beta}\wedge\phi^i-\phi^i\wedge\bpsi^{i\alpha}\sigma_{\mu\alpha\beta}\bar{\eps}^\beta)~,\\
  \delta \phi^i&\ =\ \sqrt{2}\eps\psi^i~,\\
  \delta \psi^{i \alpha}&\ =\ -4\sqrt{2}\alpha\eps^\alpha \eps^{jkli}[\phi^j,\phi^k,\phi^l]+\sqrt{2}\di (\sigma^\mu\bar{\eps})^\alpha \nabla_\mu \phi^i+\frac{\di}{\sqrt{2}\kappa}[\phi^j,\bphi^j,\phi^i]\eps^\alpha~.
\end{aligned}
\end{equation}

As stated above, supersymmetry is enhanced to $\CN=8$ for the choice $\alpha=\frac{\di}{4!\kappa}$. The additional supersymmetry transformations are parameterized by spinors $\eps^{ij}$ in the $\mathbf{6}$ of $\sSU(4)$, cf.\ \cite{Bagger:2008se}:
\begin{equation}
\begin{aligned}
 \delta\phi^i &\ =\  \sqrt{2}\bar{\eps}^{ij}\bar{\psi}_j~,\\
 \delta\bar{\psi}_i &\ =\  -\di\sqrt{2}\sigma^\mu\eps_{ij}\nabla_\mu \phi^j+[\phi^j,\phi^k,\bar{\phi}_j]\eps_{ik}+[\phi^j,\phi^k,\bar{\phi}_i]\eps_{jk}~,\\
 \delta A_\mu &\ =\ -\di \eps_{ij}\sigma_\mu \phi^i\wedge \psi^j+\di \bar{\eps}^{ij}\sigma_\mu \bar{\phi}_i\wedge \bar{\psi}_j~.
\end{aligned}
\end{equation}

\subsection{General 3-algebras}

If $\CA$ is a 3-Lie algebra, the 3-product is totally antisymmetric and the tensor $t_{ijkl}$ in \eqref{superpotential} becomes proportional to $\eps_{ijkl}$. Therefore, the theories arising from the action $S=S_0+S_1$ have $\CN=8$ for $\alpha=\frac{\di}{4!\kappa}$ or $\CN=2$ at generic values of $\alpha$. If $\CA$ is not a 3-Lie algebra, but one of the generalized 3-algebras discussed in Section 3, its 3-product is not totally antisymmetric and allows for a much more general tensor $t_{ijkl}$, containing e.g.\ terms of the form $\eps_{ab}\eps_{cd}$, where $a,b=1,2$ and $c,d=3,4$, which break the flavor symmetry group.

Note that in the Hermitian case, both the 3-bracket $[\cdot,\cdot\,;\,\cdot]$ and the scalar product $(\cdot,\cdot)_\CA$ have antilinear slots and one therefore has to employ also {\em anti}-chiral superfields when writing down the chiral superpotential terms. Moreover, to reproduce the interaction terms of the $\CN=6$ model of \cite{Bagger:2008se} by using $\CN=2$ superfields, one has to choose an $\sU(1)\cong \sSO(2)$ action within the $\sSU(4)$ R-symmetry, which becomes the R-symmetry of the $\CN=2$ superfields. This choice determines the supersymmetries which will be realized linearly, and also requires the introduction of new superfields $\Xi^i$, which are defined as linear combinations of the original chiral superfields $\Phi^i$. As the principle is clear, we refrain from going into further detail and turn our attention towards higher manifest supersymmetry.

\section{$\CN=4$ superfield formulation}

In this section, we will formulate a BLG-like action with manifest $\CN=4$ supersymmetry. There are two superspace approaches to theories with higher supersymmetries: harmonic superspace, which has recently been used in \cite{Buchbinder:2008vi} for studying the ABJM model, and projective superspace \cite{Ketov:1987yw,Lindstrom:1987ks,Lindstrom:1989ne,GonzalezRey:1997db}, which we employ here.  For a detailed discussion of superconformal field theories in projective superspace, see \cite{Kuzenko:2007qy}.

Note that a number of $\CN=4$ supersymmetric Chern-Simons matter theories have been constructed in the context of certain Janus configurations of $d=4$, $\CN=4$ super Yang-Mills theory \cite{Gaiotto:2008sd} by Gaiotto and Witten (GW). The GW models have matter fields in a different representation of the gauge
group compared to our situation and in particular, these theories do not contain the BLG model; in \cite{Hosomichi:2008jd}, however, they have been extended by twisted hypermultiplets to reproduce the BLG model. Other extensions reproduce the ABJM model and variants thereof \cite{Hosomichi:2008jb}.

\subsection{Three-dimensional projective superspace}

The $\CN=2$ superfield description of the BLG model was obtained from a dimensional reduction of four-dimensional $\CN=1$ superspace to three dimensions, and we expect the same to hold true for the $\CN=4$ description in terms of projective superspace. We thus start from the space $\FR^{1,2|8}$ with Gra\ss mann-odd derivative operators $D_{i\alpha}$ and $\bar{D}^i_\ald$ satisfying the algebra
\begin{equation}
\{D_{i\alpha},D_{j\beta}\}\ =\ 0~,~~~\{\bar{D}^i_{\ald},D^j_{\bed}\}\ =\ 0~,~~~\{D_{i\alpha},\bar{D}^j_\ald\}\ =\ -2\di\delta^j_i\sigma^\mu_{\alpha\ald}\dpar_\mu~.
\end{equation}
We now extend this superspace by the auxiliary space $\CPP^1$, which we parameterize by homogeneous coordinates $\lambda^i\in \FC^2\setminus (0,0)$. We also introduce the shorthand notation $\lambda_i:=\eps_{ij}\lambda^j$, i.e.\ $(\lambda_1,\lambda_2)=(-\lambda^2,\lambda^1)$. On $\FR^{1,2|8}\times \CPP^1$, define the following elements of $T\FR^{1,2|8}\otimes \CO(1)$:
\begin{equation}
 \nabla_\lambda\ =\ \lambda^iD_i\eand \bar{\nabla}_\lambda\ =\ \lambda_i\bar{D}^i~.
\end{equation}
Note that $\nabla_\lambda$ and $\bar{\nabla}_\lambda$ generate a $(0|2)$-dimensional Abelian subgroup parameterized by $\lambda$ inside a $(0|4)$-dimensional Abelian subgroup. Dividing $\FR^{1,2|8}\times \CPP^1$ by the latter, we obtain projective superspace $\FS$. That is, functions on $\FS$ are annihilated by $\nabla_\lambda$ and $\bar{\nabla}_\lambda$ for all $\lambda$. 

An $\sSU(2)$-invariant measure on $\FS$ can be defined after introducing an arbitrary `dual coordinate' $\pi_i$ on $\CPP^1$ satisfying the constraint $\pi_i\lambda^i=2$: This measure is given by the differential operators $\Delta=\pi^iD_i$ and $\bar{\Delta}:=\pi_i\bar{D}^i$ together with the line element $\lambda_i\dd \lambda^i$ and the volume element $\dd^3 x$:
\begin{equation}
 \mu\ :=\ \dd^3x \Delta^2\bar{\Delta}^2\lambda_i\dd \lambda^i~.
\end{equation}
Manifestly $\CN=4$ supersymmetric actions can now be written down as an integral over $\mu$ of a real functional of superfields on $\FS$.

For simplicity, we will work on the patch $\lambda^1\neq 0$ in the following and put $\lambda^i=(1,\zeta)$ as well as $\pi^i=(\frac{1}{\zeta},-1)$. The formulas then simplify to
\begin{equation}
 S\ =\ \frac{1}{2\pi\di}\int \dd^3x\, \CD^2\bar{\CD}^2\oint_\CC \frac{\dd \zeta}{\zeta} f~,
\end{equation}
where $\CD_\alpha:=D_{1\alpha}$, $f$ is a real function defined on an open subset $U\subset\FS$ and $\CC$ is a contour within $U$ avoiding any poles of $f$.

\subsection{Field content}

The matter field content is encoded in two real $\CO(2p)$-multiplets with $p=2$. Such a multiplet corresponds to a hypermultiplet in four dimensions and contains the degrees of freedom of two chiral multiplets in $\CN=2$ superspace language. Its expansion reads as\footnote{Strictly speaking, the expression in \eqref{Eq:O4} is for the ratio of a proper $\CO(4)$ multiplet and a chosen section of $\CO(4)$ given by $\zeta^2$. Thus, the degrees of $\zeta$ in our convention for $\CO(2p)$-multiplets are shifted compared to those used in \cite{Lindstrom:1989ne}.}
\begin{equation}\label{Eq:O4}
 \eta\ =\ \bar{\Phi}\frac{1}{\zeta^2}+ \bar{\Sigma}\frac{1}{\zeta}+X-\zeta\Sigma+\zeta^2\Phi~,
\end{equation}
where $X=\bar{X}$. The real structure acts by conjugation and by sending $\zeta\mapsto -(\bar{\zeta})^{-1}$. From the condition that these fields are supported on $\FS$ and thus are annihilated by $\nabla_\lambda$ and $\bar{\nabla}_\lambda$, one concludes that $\Phi$ is a chiral superfield, $\Sigma$ is a complex linear superfield $\bar{D}^2\Sigma=0$ and $X$ is a real unconstrained superfield, which will turn out to be purely auxiliary. The complex linear superfield appears as the dual of another chiral superfield in the action.

For the gauge field, we use a {\em tropical multiplet} $\CV$, which is defined around the equator of the auxiliary $\CPP^1$:
\begin{equation}
 \CV(\zeta,\bar{\zeta})\ =\ \sum_{n=-\infty}^\infty v_n\zeta^n~
\end{equation}
with components $v_{-n}=(-1)^n\bar{v}_n$. There are various gauge choices one can impose, cf.\ e.g.\ \cite{Lindstrom:1989ne}. First of all, we can gauge away all degrees $v_{\pm n}$ with $n\geq 2$, which makes the tropical multiplet look similar to that of a shifted, real $\CO(2)$ multiplet. We furthermore switch to a Lindstr{\"o}m-Ro\v{c}ek gauge, in which only the (anti-)chiral piece of $v_{-1}$ ($v_1$) survives. In this gauge, $v_0$ is a real superfield, playing the role of the $\CN=2$ vector superfield $V$, and $v_{\pm1}$ are its superpartners, i.e.\ $v_1$ is a chiral superfield and $v_{-1}$ its complex conjugate. 

As before, we assume that the two $\CO(4)$ multiplets $\eta_{1,2}$ take values in a 3-algebra $\CA$ (or its complexification if $\CA$ is real), while the tropical multiplet takes values in the associated Lie algebra $\frg(\CA)$.

\subsection{The superfield action}

The field content and the form of \eqref{CSaction1} lead us to an $\CN=4$ action:
\begin{equation} \label{CSaction2}
 S\ =\ \int \mu ~\kappa \Big(\di ( \CV,(\bar{\CD}_{\alpha}\CD^\alpha \CV))_\frg+\tfrac{2}{3}(\CV,\{(\bar{\CD}^\alpha \CV),(\CD_\alpha \CV)\})_\frg\Big)+\left(\bar{\eta}_k,\de^{2\di\CV}\eta_k\right)_\CA~,
\end{equation}
where, as above, $\CD_\alpha=D_{1\alpha}$. Let us decompose this action into $\CN=2$ superfields to demonstrate its relation with \eqref{CSaction1}. First, we examine the Chern-Simons part. In this part, the chiral derivatives $\CD$ and $\bar{\CD}$ together with the total antisymmetry of the cubic term in $\CV$ annihilate the chiral superfield in the tropical multiplet, which leaves us with terms containing nothing but the $\CN=2$ vector superfield $v_0$. To see this, recall that in three dimensions $\{\CD_\alpha,\bar{\CD}^\alpha\}=\eps^{\alpha\beta}\{\CD_\alpha,\bar{\CD}_\beta\}=-2\di\eps^{\alpha\beta}\sigma^\mu_{\alpha\beta}\dpar_\mu=0$, as the $\sigma$-matrices can be chosen to be symmetric. Thus the first term $ (\CV,(\bar{\CD}_{\alpha}\CD^\alpha \CV))_\frg$ reduces in Lindstr{\"o}m-Ro\v{c}ek gauge to $( v_0,(\bar{\CD}_{\alpha}\CD^\alpha v_0))_\frg$ under the integral. The terms which remain of  $\tfrac{2}{3}(\CV,\{(\bar{\CD}^\alpha \CV),(\CD_\alpha \CV)\})_\frg$, after the contour integration in Lindstr{\"o}m-Ro\v{c}ek gauge is performed, and which contain fields besides $v_0$ are
\begin{equation}
(v_{-1},\{\CD^\alpha v_0,\bar{\CD}_\alpha v_1\})_\frg+(v_{0},\{\CD^\alpha v_{-1},\bar{\CD}_\alpha v_1\})_\frg+(v_{1},\{\CD^\alpha v_{-1},\bar{\CD}_\alpha v_0\})_\frg~.
\end{equation}
Using partial integration as well as cyclicity of $(\cdot,\cdot)_\frg$, one can rewrite this expression so that it is proportional to  $\{\CD^\alpha,\bar{\CD}_\alpha\}$, which is identically zero. Thus, the $\CN=4$ Chern-Simons part reduces to the $\CN=2$ Chern-Simons part.

The part containing the hypermultiplets requires a lengthier analysis; we can essentially follow the discussions of \cite{Lindstrom:1989ne,GonzalezRey:1997qh,GonzalezRey:1997xp}. We split the exponential of the gauge potential according to
\begin{equation}\label{Eq:RH}
 \de^{2\di \CV(\zeta)}\ =\ \de^{2\di \CV_-(\zeta)}\de^{2\di \CV_+(\zeta)}~,
\end{equation}
where\footnote{Note that $v_i=\CV_i$ only in the Abelian case.} 
\begin{equation}
 \CV_-(\zeta)\ =\ \sum_{i=1}^\infty \bar{\CV}_i\left(-\frac{1}{\zeta}\right)^i~,~~~\CV_+(\zeta)\ =\ \sum_{i=0}^\infty \CV_i\zeta^i~,
\end{equation}
and introduce gauge covariant spinor derivatives
\begin{equation} \label{def:CovariantSpinorDerivatives}
 \tilde{\nabla}\ :=\ \de^{2\di \CV_+}\nabla \de^{-2\di \CV_+}\ =\ \de^{-2\di \CV_-}\nabla \de^{2\di \CV_-}\eand\tilde{\bar{\nabla}}\ :=\ \de^{2\di \CV_+}\nabla \de^{-2\di \CV_+}\ =\ \de^{-2\di \CV_-}\bar{\nabla}\de^{2\di \CV_-}~,
\end{equation}
where the equalities hold due to $\nabla \CV=\bar{\nabla} \CV=0$. These derivatives define covariant projective superfields
\begin{equation}
 \tilde{\eta}_k\ =\ \de^{2\di \CV_+}\eta_k~,~~~\tilde{\bar{\eta}}_k\ =\ \bar{\eta}_k\de^{2\di \CV_-}~,
\end{equation}
which are annihilated by the covariant derivatives \eqref{def:CovariantSpinorDerivatives}. Note that one has to take the appropriate actions of these derivatives containing terms in $\frg(\CA)$ on the real $\CO(4)$ multiplets $\tilde{\eta}_k$ and $\tilde{\bar{\eta}}_k$, which are $\CA$-valued.

Introducing the derivatives \eqref{def:CovariantSpinorDerivatives} effectively generates central charges in the $\CN=2$ spinor derivatives \cite{GonzalezRey:1997xp}:
\begin{equation}
\tilde{\nabla}\ =\ \tilde{D}_1+\zeta \tilde{D}_2~,~~~\{\tilde{D}^{1\alpha},\tilde{D}_{2\alpha}\}=:4\eps_{\alpha\beta}\CW~,
\end{equation}
which implies the covariant complex linearity condition $(D_1)^2\tilde{\Sigma}=2 \CW \tilde{\Phi}$ for the covariant $\CO(4)$ multiplet. The matter part of the action can now be rewritten as
\begin{equation}
S_{m}\ =\ \int\mu\left(\tilde{\bar{\eta}}_k,\tilde{\eta}_k\right)\ :=\ \int \dd^3x\int \dd^4\theta\Big((\tilde{\bar{\Phi}},\tilde{\Phi})_\CA-(\tilde{\bar{\Sigma}},\tilde{\Sigma})_\CA+(\tilde{\bar{X}},\tilde{X})_\CA\Big)~,
\end{equation}
where $\Phi$ is again chiral, $\Sigma$ is (covariantly) complex linear, and we suppressed the flavor index $k$. This action can be obtained from \cite{GonzalezRey:1997qh}
\begin{equation}
\begin{aligned}
S_{m}'\ =\ \int \dd^3x\int \dd^4\theta\Big(&(\tilde{\bar{\Phi}},\tilde{\Phi})_\CA-(\tilde{\bar{\Sigma}},\tilde{\Sigma})_\CA+(\tilde{\bar{X}},\tilde{X})_\CA\\
&+(Y,((\bar{D}_1)^2\tilde{\bar{\Sigma}}+2\CW\tilde{\bar{\Phi}}))_\CA+(\bar{Y},((D_1)^2\tilde{\Sigma}-2\bar{\CW}\tilde{\Phi}))_\CA\Big)
\end{aligned}
\end{equation}
with an unconstrained field $\tilde{\Sigma}$ by integrating out the fields $Y,\bar{Y}$. The auxiliary field $X$ decouples and, after integrating out the fields $\tilde{\Sigma}$ and $\tilde{\bar{\Sigma}}$, a second covariant chiral superfield $\tilde{\Psi}:=(D_1)^2Y$ appears:
\begin{equation}
\begin{aligned}
S_{m}''\ =\ \int \dd^3x\int \dd^4\theta\Big(&(\tilde{\bar{\Phi}},\tilde{\Phi})_\CA-(\tilde{\bar{\Psi}},\tilde{\Psi})_\CA+(\tilde{\bar{X}},\tilde{X})_\CA\Big)\\
&+\int \dd^3x \int \dd^2\theta (\tilde{\Psi},-2\CW\tilde{\bar{\Phi}})_\CA+\int \dd^3x \int \dd^2\bar{\theta} (\tilde{\bar{\Psi}},2\bar{\CW}\tilde{\Phi})_\CA~.
\end{aligned}
\end{equation}
The second line is responsible for generating the usual mass-coupling superpotential terms in the description of a hypermultiplet in terms of two $\CN=2$ chiral superfields.

A detailed analysis of the $\CN=1$ components of the interaction terms would essentially require that one solves the Riemann-Hilbert problem (\ref{Eq:RH}): splitting $\CV$ into $\CV_+$ and $\CV_-$. Alternatively, one can consider the symmetries of the action. Supersymmetry requires integration over the full measure $\mu$. Demanding classical conformal invariance as well as no higher derivatives reduces the choices of the matter-field action to $S_m$. As the actual BLG model has to be included in the set of $\CN=4$ BLG-type models, we conclude that \eqref{CSaction2} is in fact the BLG model, if $\CA$ is a 3-Lie algebra. In other cases, the interaction terms will differ and $\CN=4$ or $\CN=6$ supersymmetry will be realized.

\section{Conclusions}

We have presented a complete classification of functorial representations on $*$-algebras of the generalized metric 3-algebra structures exposed in \cite{deMedeiros:2008zh}. We found the four families \eqref{eqA} of such representations in the real case, while the Hermitian case allowed only for the single one \eqref{eqB}. These altogether five classes of 3-brackets give a wealth of explicitly realized matrix 3-algebras of arbitrary rank. What is more important, these brackets may be reasonable candidates for 3-algebraic structures on matrices which are rich enough to represent any generalized metric 3-algebra. We hope to address this issue in more detail elsewhere.

We have also given manifestly $\CN=2$ and $\CN=4$ invariant formulations of BLG-type models based on these real and Hermitian 3-algebras. For the former, we used the dimensional reduction of ordinary $\CN=1$ superspace from four to three dimensions, while for the latter, we employed the dimensional reduction of $\CN=2$ projective superspace. The $\CN=4$ formulation of Chern-Simons matter theories is completely new, and might find further applications besides the description of multiple M2-branes. Both the $\CN=2$ and the $\CN=4$ formulations can be endowed with any of the generalized metric 3-algebra structures in our classification.  In general, however, such actions have less than the $\CN=8$ supersymmetry of the original BLG theory that is based on the 3-Lie algebra $A_4$. Note that similarly, manifestly $\CN=4$ supersymmetric formulations of the ABJM and the GW models can also be given.

One might consider the absence of any explicit 3-product in the $\CN=4$ formulation as evidence that the more general 3-algebras appearing are not fundamental, but merely appear in a rewriting of the theory. This would suggest that this 3-algebra rewriting does not provide any deeper insight than its equivalent gauge theory formulation.   One can contemplate, however, that the presence of the underlying 3-algebra structure does indicate some hidden symmetries and conserved currents.  We hope that our actions and 3-algebra classification can shed some light on this issue.  Also, the fact that 3-Lie algebras neatly fit into the framework of strong homotopy Lie algebras \cite{Lazaroiu:2008aa} hints at a more subtle role of the 3-algebras. While 3-Lie algebras satisfy the homotopy Jacobi identities of strong homotopy Lie algebras, the generalized 3-Lie algebras satisfy the strong homotopy pre-Lie algebra identities. All this suggests that generalized 3-Lie algebras may provide some new insight into the properties of the theories that we have presented here.

\acknowledgements
SCh is grateful to Martin Ro\v{c}ek and Ulf Lindstr\"om for useful conversations. VD wishes to thank Nicolas Thiery for providing a \texttt{MuPAD} module to handle $*$-algebras. SCh  is supported by the Science Foundation Ireland Grant No.\ 06/RFP/MAT050 and by the European Commission FP6 program MRTN-CT-2004-005104. VD and CS are supported by an IRCSET (Irish Research Council for Science, Engineering and Technology) postdoctoral fellowship.

\appendices

\subsection{De Wit-Freedman transformation}

After performing a supersymmetry transformation on a supervector field in Wess-Zumino gauge, one needs to perform an additional gauge transformation in order to return to this gauge. This will also affect the matter superfields, and we collect all the relevant formulas here for reference. We assume that the chiral superfield is in the fundamental representation. Also, contrary to the conventions of \cite{Wess:1992cp}, we assume anti-Hermitian generators, so that
\begin{equation}
 F_{\mu\nu}\ :=\ \dpar_\mu A_\nu-\dpar_\nu A_\mu+[A_\mu,A_\nu]~,~~~\nabla_\mu\Phi\ =\ \dpar_\mu\Phi+A_\mu\Phi~,~~~\nabla_\mu\bar{\Phi}\ =\ \dpar_\mu\bar{\Phi}-\bar{\Phi} A_\mu~.
\end{equation}
As gauge transformations, we have
\begin{equation}
 \de^{2\di V'}\ =\ \de^{2\Lambda^\dagger}\de^{-2\di V}\de^{-2\Lambda}~,~~~\Phi'\ =\ \de^{2\Lambda}\Phi~,~~~\bar{\Phi}\ =\ \bar{\Phi}\de^{-2\Lambda^\dagger}~,
\end{equation}
where we suppressed the flavor index on the matter fields. Here $\Lambda$ is a chiral superfield, and the choice of $\Lambda$ taking the supersymmetry variation 
\begin{equation}
 \delta V\ :=\ \eps^\alpha Q_\alpha V-\bar{\eps}^\ald Q_\ald V
\end{equation}
back to Wess-Zumino gauge reads in chiral coordinates $y$ as
\begin{equation}
 \Lambda(y)\ =\ \di (\theta\sigma^{\hat{\mu}}\bar{\eps}) A_{\hat{\mu}}(y)+\theta^2\bar{\eps}\bar{\lambda}(y)~.
\end{equation}
To shorten the expressions, we used four-dimensional indices, which are marked by a hat. In components, the combined supersymmetry and gauge transformations have the following effect:
\begin{equation}
\begin{aligned}
  \delta \phi&\ =\ \sqrt{2}\eps\psi~,\\
  \delta \psi^\alpha&\ =\ \sqrt{2}\eps^\alpha F+\sqrt{2}\di (\sigma^{\hat{\mu}}\bar{\eps})^\alpha \nabla_{\hat{\mu}} \phi~,\\
  \delta F&\ =\ -\sqrt{2}\di ((\nabla_\mu\psi)\sigma^\mu\bar{\eps})+2(\bar{\eps}\bar{\lambda})\phi~.
\end{aligned}
\end{equation}
For the chiral superfield, and the vector superfield transformation are covariantized:
\begin{equation}
\begin{aligned}
  \delta A_{\hat{\mu}} &\ =\ \di(\eps\sigma_{\hat{\mu}}\bar{\lambda}-\lambda\sigma_{\hat{\mu}}\bar{\eps})~,\\
  \delta \lambda^\alpha \ =\ (\sigma^{\hat{\mu}\hat{\nu}})^\alpha{}_\beta \eps^\beta F_{\hat{\mu}\hat{\nu}}+&\di\eps^\alpha D~,~~~
\delta \bar{\lambda}^\ald \ =\ (\bar{\sigma}^{\hat{\mu}\hat{\nu}})^\ald{}_\bed \bar{\eps}^\bed F_{\hat{\mu}\hat{\nu}}-\di\eps^\ald D~,\\
  \delta D&\ =\ -(\nabla_{\hat{\mu}}\lambda)\sigma^{\hat{\mu}}\bar{\eps}-\eps\sigma^{\hat{\mu}}(\nabla_{\hat{\mu}}\bar{\lambda})~.
\end{aligned}
\end{equation}

\end{document}